# AN INFORMATION-BASED MODEL SELECTION CRITERION FOR DATA-DRIVEN MODEL DISCOVERY

**Michael C. Chung**
Division of Chemical Biology and Medicinal Chemistry
College of Pharmacy
University of Texas at Austin; Austin, TX 78712

**Alen Zacharia**
Department of Physics
Carnegie Mellon University; Pittsburgh, PA, 15213

**Juan Guan**
Division of Chemical Biology and Medicinal Chemistry
College of Pharmacy
University of Texas at Austin; Austin, TX 78712
juanguan@utexas.edu

## ABSTRACT

Data-driven model discovery (DDMD) algorithms are powerful tools for extracting interpretable symbolic models from data. However, identifying the model that best balances goodness-of-fit and sparsity is often a laborious process requiring user fine-tuning, is prone to overfitting, and results may significantly vary depending on model initialization and specific training procedure. Here, we present a sparse regression algorithm that automatically and adaptively generates candidate models, and uses a novel sample-length-scaling logarithmic information criterion (SLIC) to identify the best model from these candidates. We demonstrate that SLIC greatly outperforms other popular information criteria in extracting the correct model from the data of several nonlinear ordinary and partial differential equations. Then, we demonstrate SLIC's ability to discover interpretable models from experimental datasets in fluid dynamics and nanotechnology that generate new testable predictions.

## 1 Introduction

Models, which we loosely define as mathematical expressions that underpin relationships within data, lie at the heart of all science and engineering disciplines. Unfortunately, manually constructing a model for a system requires extensive mathematical and domain expertise. With the increasing capacity to collect large datasets and ever-growing computing power, we have seen the emergence of data-driven model discovery (DDMD) algorithms to address this challenge. This paradigm shift departs from traditional manual construction of models and aims at extracting accurate, interpretable models of complex systems directly from data. These techniques are numerous and include dynamic mode decomposition and Koopman methods Schmid, Budišic´ et al., Wilson, Williams et al., symbolic regression Cranmer, Reinbold et al., Udrescu and Tegmark, and regression-based library/dictionary methods Brunton et al., Brunton and Kutz, Hastie et al.. However, discovering the best overall model that is both accurate and sparse (i.e. possessing the minimal number of mathematical terms possible) remains a central challenge in DDMD.

The major challenge in DDMD entails finding the optimally sparse model within a given model architecture. This requires both (i) determining the optimal sparsity-enforcing parameters and (ii) robust model selection. The optimal sparsity-enforcing parameters, which in this work are pruning thresholds that remove unnecessary terms from a model, are often unknown a priori. The choice of such parameters is a delicate problem: if sparsity-enforcement is too lax, the model would be grossly overfit with unnecessary terms; if too aggressive, the model would not capture the full scope of the observed phenomena. Thus, setting the optimal sparsity-enforcing thresholds often requires extensive user-specified input and fine-tuning which cannot generalize easily per its case-by-case nature or tackle real-time applications due to laborious and time-consuming user interventions. As for the second requirement, though much effort has been directed at refining the procedure by which models are estimated from data Champion et al., Messenger and Bortz [a,b], Egan et al., Goyal and Benner, Zhang and Lin, Hirsh et al., Course and Nair, fundamental research on designing novel model selection criteria suitable for DDMD is lacking. Specifically, many widely used model selection criteria prioritize goodness-of-fit over sparsity, resulting in overfit models that cannot generalize or correctly predict the behavior of the system beyond training data. This tendency to overfit is due to the functional forms of these criteria, which, as a result of differences in scaling with data quantity, weigh model accuracy disproportionately more than model sparsity.

Here, to address this challenge, we present a sparse regression-based algorithm that adaptively auto-generates sparsity-enforcing parameters directly from the data, thus circumventing the need for user fine-tuning, and utilizes a novel sample-length-scaling logarithmic information criterion (SLIC) designed to robustly and automatically identify the model that best achieves both goodness-of-fit and model simplicity. The auto-thresholding and SLIC scoring work efficiently towards discovering the best overall model (Fig. 1).

We first present a concise problem statement, a brief motivation for SLIC and its derivation, and the SLIC-sparse regression algorithm. Then, we demonstrate that SLIC can robustly and automatically extract the governing equations of numerous simulated ordinary and partial differential equations. Finally, we demonstrate the utility of SLIC to extract mathematical models from real experimental datasets from fluid dynamics and nanoscience, which give accurate predictions on unseen experimental conditions.

## 2 Problem statement, our information criterion, and the SLIC-sparse regression algorithm

### 2.1 Problem Statement

Here, we provide a simple outline of the problem at hand. Our data consists of a pair of matrix inputs, $X = [x_1 \ldots x_m] \in R^{n \times m}$, and corresponding outputs, $Y = [y_1 \ldots y_p] \in R^{n \times p}$, connected by an unknown mathematical relationship. Here, n denotes the number of observations, m denotes the number of state variables (features), and p the number of outputs (targets). For example, for an ODE describing a dynamical system, each column of $X$ contains a measured state variable over its time course and each column of $Y$ contains the corresponding numerically estimated time derivatives of the state variables. For time-dependent PDEs, n is simply the total number of spatiotemporal observations, which scales multiplicatively with the number of dimensions.

Here, we assume that the mathematical relationship can be modeled using a nonlinear regression framework Brunton et al., Hastie et al.: $Y = \Theta(X)\Xi + \eta$, where $\Theta(X) \in R^{n \times l}$ is a user-specified matrix of nonlinear transformations of the inputs, $\Xi \in R^{l \times p}$ is a matrix of unknown model coefficients, and $\eta$ is zero-mean, isotropic Gaussian noise with unknown noise strength; l is the number of library terms, which may include polynomial nonlinearities, trigonometric terms, etc.

To find $\Xi$, we solve the following optimization problem:

$$\hat{\Xi} = \arg\min_{\Xi} \|Y - \Theta(X)\Xi\|_2^2 + \lambda^2 R(\Xi) \tag{1}$$

Where $\|\cdot\|_2^2$ is the sum of square errors, $R(\cdot)$ is a regularization term, and $\lambda$ is the concomitant regularization coefficient. In this work, we take $R(\cdot)$ to be the $L_0$ norm, which penalizes the number of nonzero coefficients in the number. In this case, $\lambda$ manifests itself as a thresholding parameter that effectively prunes terms in $\Xi$ whose magnitude falls beneath $\lambda$. It is thus imperative to choose $\lambda$ wisely. Our algorithm for solving this problem without having to specify $\lambda$ is detailed in Section 2.3.

An important special case considered in a majority of this study is when we want to extract governing equations from timeseries data, i.e. $Y = \dot{X}$. To avoid taking numerical derivatives, we cast the problem into its weak form, as detailed in Materials and Methods 5.45.4.2.

### 2.2 An information-based model selection criterion for model discovery

Many principled methods exist for model selection Ding et al. [a]. Ideal model selection will achieve both simplicity and accuracy in the returned model. One important, widely used class of model selection techniques are known as information-based model selection criteria, which aim to minimize metrics related to the information content of probability distributions, such as the Kullback-Leibler divergence (KLD) Stoica and Selen.

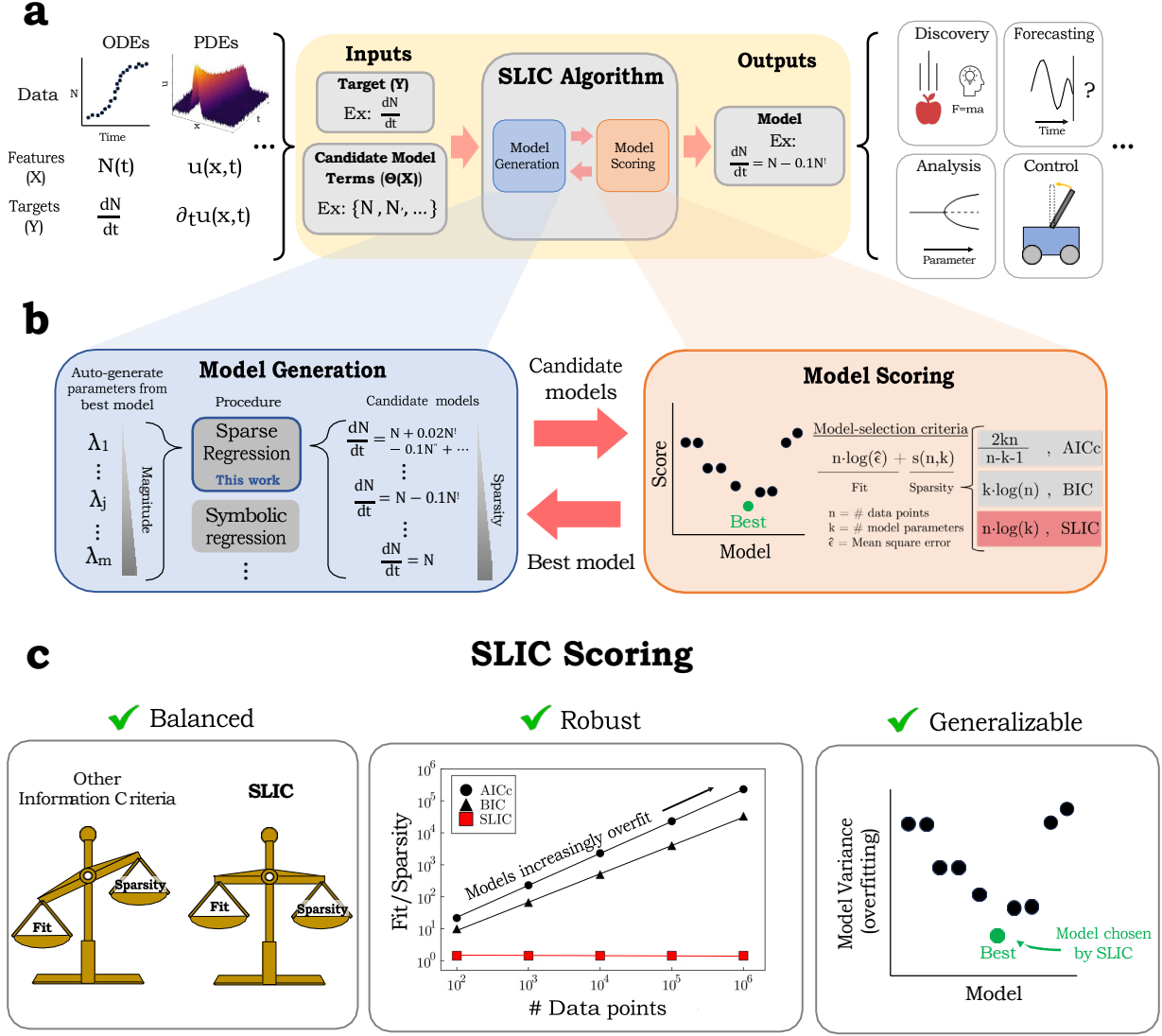

Figure 1: (a) The overview of the SLIC algorithm. We want to extract models (governing equations, etc.) from data, which can originate from diverse sources. To do so, we use our iterative model generation-model scoring loop to optimally prune input models, which are then updated. The output model can be used for myriad purposes, including scientific discovery, forecasting, analysis, and control. (b) Overview of the central iterative component of the algorithm: adaptive, automatic extraction of sparsity-enforcing parameters from input data are used to generate candidate models ("Model Generation"); and our sample-length-scaling logarithmic information criterion (SLIC) designed to avoid overfitting when selecting best models from candidate models ("Model Scoring"). The resulting best model determined by SLIC is then used to regenerate sparsity-enforcing parameters on the next iteration (see text and Methods for details). (c) Benefits of SLIC scoring. Left: SLIC equitably weights fit and sparsity. Center: The y-axis is the fit-to-sparsity ratio, which serves as a proxy for quantifying overfitting. Right: SLIC is equivalent to minimizing model variance.

Generally, many information criteria (denoted as IC below) take the form Stoica and Selen:

$$\text{IC} \approx -2\log P(D|\hat{\theta}) + s(n, k) \quad (2)$$

Where n is the number of data points, k is the model complexity (i.e. the number of parameters), $P(D|\hat{\theta})$ is the likelihood evaluated at the most likely parameters (denoted as $\hat{\theta}$) for a given model, and s(n, k) is a function that penalizes model complexity. Under the further assumption that this likelihood is univariate Gaussian, we get the following form of these scoring criteria Stoica and Selen:

$$IC(\hat{\epsilon}, k) = n \log(\hat{\epsilon}) + s(n, k) \tag{3}$$

Where $\hat{\epsilon}$ is the mean-squared error of the model prediction, which is also the maximum likelihood estimation of the unknown noise strength.

Different model selection criteria differ primarily in their choice of $s(n, k)$. For example, for the Akaike Information Criterion (AIC) Akaike, $s_{AIC}(n, k) = 2k$; for the Bayesian Information Criterion (BIC) Schwarz, $s_{BIC}(n, k) = k \log(n)$. The model with the minimum information criterion score is chosen and presumed to be the model that best balances goodness-of-fit with the model complexity. Thus, depending on the functional form of $s(n, k)$ different information criteria may select different models.

Interestingly, a common feature of these information criteria taking the form (3) is that as n grows large, the sparsity-enforcing component becomes negligible in comparison to the goodness-of-fit term that scales linearly with n (i.e. $\lim_{n\to\infty} n \log(\hat{\epsilon})/s(n, k) \to \infty$). Although model complexity may increase with the amount of collected data in certain applications, it is clearly not the case when the underlying model should be invariant to the amount of collected data. To combat such issues, users are required to resort to more complex procedures such as fine-tuning of sparsity-enforcing parameters, data subsampling, etc.

With this in mind, we sought an improved $s(n, k)$, and hence an information-based scoring criterion, suitable for the task of model selection in DDMD. Here we give an argument for such a criterion based on scaling and symmetry principles.

First, given the supposed invariance of the correct model with respect to collected data, one may suppose that $s(n, k)$ should scale asymptotically with n, i.e. $s(n, k) \sim n$. Otherwise, we would run into the same scaling issues facing other information criteria. Using this assumption, without loss of generality, the simplest functional form would be $s(n, k) = n \log(f(k))$, where $f(k)$ is a pure function of the number of free parameters. In turn, this implies $IC(\hat{\epsilon}, k) = n \log(\hat{\epsilon} f(k))$.

Next, we demand equitable penalization of goodness-of-fit and complexity. One such way to enforce this is to demand that the information criterion be symmetric in its arguments: $IC(\hat{\epsilon}, k) = IC(k, \hat{\epsilon})$. This implies:

$$\hat{\epsilon} f(k) = k \mathbf{f}(\hat{\epsilon}) \tag{4}$$

$$\frac{\mathbf{f}(\hat{\epsilon})}{\hat{\epsilon}} = \frac{f(k)}{k} \tag{5}$$

Assuming the approximate independence of $\hat{\epsilon}$ and k, the only way to satisfy this functional equation is for the ratio to be a constant, $f(k) \propto k$. We drop the superfluous proportionality factor, as only differences in these scores matter.

As this information criterion scales with sample length and possesses a logarithmic form in k, we dub it SLIC, for sample-length-scaling logarithmic information criterion:

$$SLIC(\hat{\epsilon}, k) = n \log(\hat{\epsilon} k) \tag{6}$$

Interestingly, in the sparse regression framework, where we consider our outputs to be linear combinations of nonlinear transformations of our input data, choosing a model that minimizes this information criterion score is intimately tied to choosing a model that minimizes model variance in the famous bias-variance tradeoff Hastie et al.:

$$\text{Model Variance} = \frac{\hat{\epsilon} k}{n} \tag{7}$$

Minimizing the SLIC score in (6) thus minimizes the model variance, at least for a single target variable. In this sense, models selected by SLIC will be least likely to overfit and most likely to generalize beyond the training data (See Supporting Information for a derivation of (7)). For targets with multiple variables, SLIC minimization approximates minimizing total variance ($\propto \sum_j \hat{\epsilon}_j k_j$) if deviations in total error are negligible $\sum_j \hat{\epsilon}_j k_j \approx \hat{\epsilon} \sum_j k_j = \hat{\epsilon} k$.

Given its favorable scaling with data quantity, SLIC circumvents the paradoxical issue of the increasing likelihood of overfitting with data quantity that is manifest in information criteria such as AIC and BIC. Thus, by design, SLIC equitably weights accuracy and sparsity, is robust to data quantity, minimizes user intervention, and, due to its relation with variance minimization, extracts generalizable models (Fig. 1c).

### 2.3 The SLIC-sparse regression algorithm

Here, in Algorithm 1, we give a brief description of the central component of the automatic pruning algorithm using SLIC, depicted in Fig. 1. Below, X,Y represent the input and output data, respectively; Ξ is the matrix of unknown model terms. The SLIC score of the model is SLIC($\Xi$) = n log($\hat{\epsilon}^k$), where n is the number of observations, $\hat{\epsilon}$ is the mean-squared error between the model prediction and data, and k is the model complexity, or the number of nonzero parameter, which includes the unknown noise strength.

First, we estimate the nonzero parameters by minimizing $\|Y - \Theta(X)\Xi\|_2^2$ which is easily solved via taking the pseudoinverse of $\Theta(X)$, $\hat{\Xi} = \Theta(X)^\dagger Y$. This model is then scored using SLIC. Then, an array of candidate thresholds, $\lambda$s, is generated from the estimate of $\Xi$ by simply using the absolute value of the elements of $\Xi$ itself. Then, for each $\lambda$ in the array $\lambda$s, the parameters of $\Xi$ whose magnitudes fall beneath $\lambda$ are pruned. The nonzero parameters of the pruned model are again updated by pseudoinverse. If the SLIC score of this pruned model is smaller than the current best model (according to smallest SLIC score), then this becomes the new best model. This process is continued for ten iterations. The model with the best SLIC score is returned as output.

---

**Algorithm 1** Auto-pruning SLIC Algorithm

    Inputs: $\Theta(X)$, Y
    Outputs: $\Xi$
    $\Xi \leftarrow \Theta(X)^\dagger Y$         ▷ Get initial model estimate via pseudoinverse
    BestScore $\leftarrow$ SLIC($\Xi$)         ▷ score current best model
    $\lambda$s $\leftarrow$ nonzero elements of $\Xi$         ▷ pruning parameters
    **for** $\lambda$ in $\lambda$s **do**
        $\Xi_{new} \leftarrow$ copy($\Xi$)
        inds $\leftarrow |\Xi_{new}| < \lambda$         ▷ get 'small' terms
        $0 \leftarrow \Xi_{new}$(inds)         ▷ prune 'small' terms
        $\Xi_{new} \leftarrow$ update($\Xi_{new}$)         ▷ update nonzero parameters
        NewScore $\leftarrow$ SLIC($\Xi_{new}$)
        **if** NewScore < BestScore **then**
            BestScore $\leftarrow$ NewScore
            $\Xi \leftarrow \Xi_{new}$
        **end if**
    **end for**

---

## 3 Results

### 3.1 SLIC extracts true governing equations of ordinary and partial differential from data

We set out to gauge the ability of SLIC to extract governing equations for dynamical systems from data in a sparse regression framework, which represents output data as a linear combination of nonlinear transformations of input data Brunton et al.. We compare the performance of SLIC to robustly recover models against a host of other popular information criteria in the literature, including AIC Stoica and Selen, Akaike, AICc Mangan et al., Kaptanoglu et al., Sugiura, HQIC Hannan and Quinn, BIC Schwarz, Stoica and Selen, KIC Kashyap, Bridge Criterion (BC) Ding et al. [b]. See Methods and Materials for details concerning the information criteria benchmarked against SLIC, the definition of noise percentage, a discussion of all metrics used to quantify algorithm performance, the candidate model terms used in each system, and the general preprocessing and workflow.

We first tested the algorithm on benchmark ordinary differential equations (ODEs) commonly found in the literature Messenger and Bortz [a], Brunton et al. (Fig. 2a). We found that SLIC greatly outperforms all other information criteria in recovering the true underlying model across all noise percentages, with near perfect accuracy in the recovered model terms (Fig. 2b). SLIC rarely includes false-positive model terms in the dynamics, as opposed to the other information criteria, which consistently have a high false-positive rate (FPR) (Fig. 2c). Interestingly, this trend continues even as data is downsampled (Fig. S1), where other information criteria perform more favorably, as predicted. SLIC also returns models with a smaller scaled mean absolute error (SMAE, see Methods) (Fig. S2a), indicating SLIC returns models with coefficients close to the true model coefficient values. Further, SLIC is also robust to changes in sampling frequency (Fig. S3), trajectory length (Fig. S4), and ODE parameters (Fig. S5).

### 3.2 Discovering model for center of mass dynamics of a sloshing fluid in a rectangular tank

Across science and engineering disciplines, extensive domain knowledge and mathematics expertise have often been required to parse out models from experimental data to describe underlying dynamic processes. This constraint may hinder discovery of novel mechanisms where domain knowledge has not been accumulated sufficiently. Moreover, data

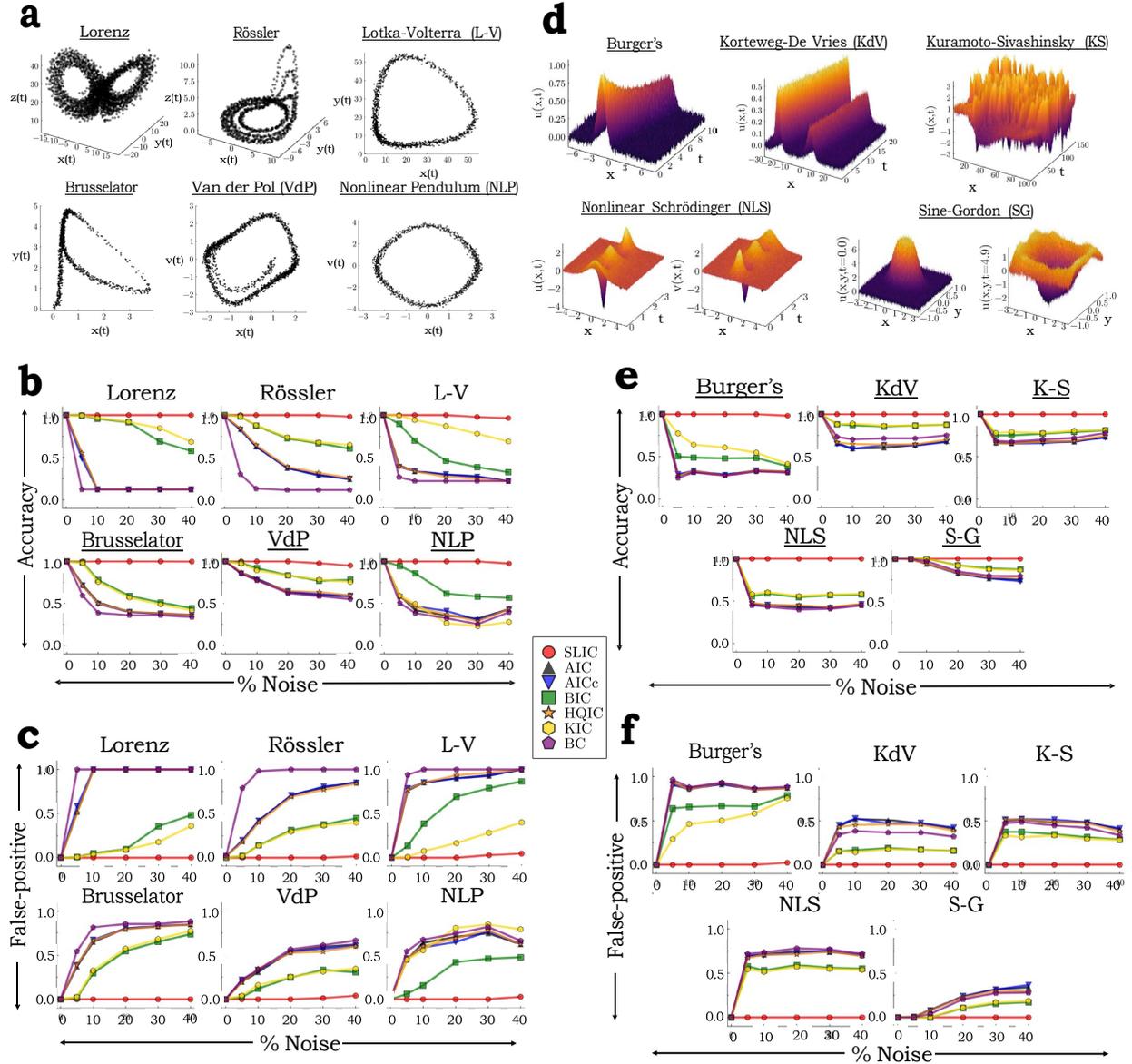

Figure 2: Performance of SLIC on ODEs and PDEs. (a,d) State-space plots of benchmark ODEs with 5% Gaussian noise (a), and PDEs with 20% Gaussian noise (d). See Methods for the equations for each system and other simulation details. (b,e) SLIC demonstrates superior recovery of the true governing equations of benchmark ODEs (b) and PDEs (e) over a range of noise levels, as opposed to other information criteria. Metric shown on y-axis is the average accuracy (n=25), which measures how often the correct model terms are identified by each model selection criterion. Accuracy ranges from 0 (correct equations are never correctly determined) to 1 (correct equations are always correctly determined). See Methods for further discussion. (c,f) SLIC rarely returns overfit models, as opposed to other information criteria. Metric shown on y-axis is the average false-positive rate (FPR) (n=25), which measures how often incorrect coefficients are identified by each model selection criterion. FPR ranges from 0 (incorrect coefficients are never included) to 1 (incorrect coefficients are always included). See Methods for further discussion.

in dynamic regimes where model predictions are the most impactful may be required as part of the training data during model construction, defeating the purpose of model predictions. In the sloshing tank example below, we test whether

SLIC can extract models without domain knowledge and make accurate predictions well beyond the training regime used during model discovery.

The dynamics of sloshing liquids is critical in disciplines in which various vessels host onboard fluids, such as aerospace engineering Abramson, Veldman et al., Hubert and fuel transport Popov et al., Romero et al.. Often, the dynamics of these onboard fluids may disrupt and possibly endanger the intended function of these various vessels. These phenomena are highly complex and typically demand expert knowledge of the underlying mechanism to understand their behavior. Thus, it is highly desirable for DDMD algorithms to aid in the discovery of the dynamics of such systems.

To experimentally assess the dynamics of sloshing fluids in a controlled manner, Ref. Bäuerlein and Avila engineered a setup in which a motor horizontally drives a rectangular tank at a specified frequency and amplitude. The images of the sloshing fluid gathered from this setup can be used to extract timeseries data of the center of mass, permitting the researchers to explore the dynamics of this system (Fig. 3a).

Here, we test if, using our algorithm, we could discover a model directly from the data in a simple manner, without any domain knowledge to fit data to a pre-determined model (i.e. not knowing the underlying dynamics or existing symmetries *a priori*). We find that with data from a single sloshing experiment without forcing, SLIC determines that the dominant dynamics is governed by a Duffing system (Fig. 3b):

$$\ddot{x} = -(7.802)^2 x - 2(0.072)\dot{x} + 0.351 x^3 \tag{8}$$

In contrast, other information criteria again return a non-sparse solution (Fig. S6a).

Next, we test if SLIC can make accurate predictions on future behaviors. First, we find that this spartan workflow can forecast quite well based on limited training data (Fig. 3b). Second, unlike the manual fitting that requires datasets from both unforced and forcing conditions, SLIC is only trained on data from the unforced condition. We therefore challenge SLIC to predict dynamics in the unseen forcing condition. Strikingly, the model extracted by SLIC from the unforced data accurately predicts the forcing behavior. SLIC prediction on the steady-state amplitude (Fig. 3c) and phase-shift (Fig. 3d) of the forced system shows excellent agreement with experimental data. Moreover, SLIC even captures the hysteretic, unstable portion of the curves where experimental data cannot be collected. This is critical, as design considerations for sloshing liquids require accurate predictions of the system in unstable regimes, where SLIC has demonstrated excellent predictive power.

### 3.3 Discovering the functional form of RNA-liposome cluster-cluster interaction rate and improving mRNA delivery in vitro based on model prediction

Lastly, we test if SLIC can perform model discovery in the face of additional unknown, model-specific parameters that must be determined via auxiliary optimization schemes. Here we test SLIC's ability to extract the aggregation model of a clinically relevant nanomaterials system Mendez-Gomez et al. [a], which requires determining an unknown exponent (the fractal dimension) via a nested optimization procedure.

Nanoscience has transformed countless disciplines in recent decades, with applications ranging from material science Malik et al. to modern therapeutics Mitchell et al.. In these applications, models informing the rational design and engineering of nanoparticle formulations is paramount for success Blanco et al., Albanese et al., Hoshyar et al.. In particular, it is often critical to understand how nanoparticles self-assemble Bassani et al., Rao et al.. One general, well-established class of models describing time-dependent self-assembly is Smoluchowski aggregation theory Elimelech et al. (see Supporting Text). Smoluchowski theory is a collection of coupled nonlinear differential equations describing how clusters of particles evolve over time:

$$\frac{dn_k}{d} = \frac{1}{2} \sum_{i+j=k} K_{ij} n_i n_j - n_k \sum_i K_i n_i \tag{9}$$

Here $n_k$ denotes 'k-mers', or clusters containing 'k' particles and $K_{ij}$ denotes the kernel function, which is symmetric and dictates the rate of interaction between clusters (Fig. 4a) and characterizes the underlying transport and aggregation mechanism of the system (Fig. 4b). The first term on the RHS of the above expression determines the ways in which one can create k-mers via the assembly of smaller clusters; the second term determines how k-mers are destroyed via merging with other clusters.

Thus, for many nanoparticle systems, the goal is to understand the nature of the kernel and its associated physics. However, manual trial-and-error kernel fitting procedures are time-consuming and impractical, especially if the assembly mechanism is not known. Thus, we reason that DDMD can be an extremely powerful approach for understanding the dominant physics of cluster-cluster interactions by extracting the functional form of the kernel.

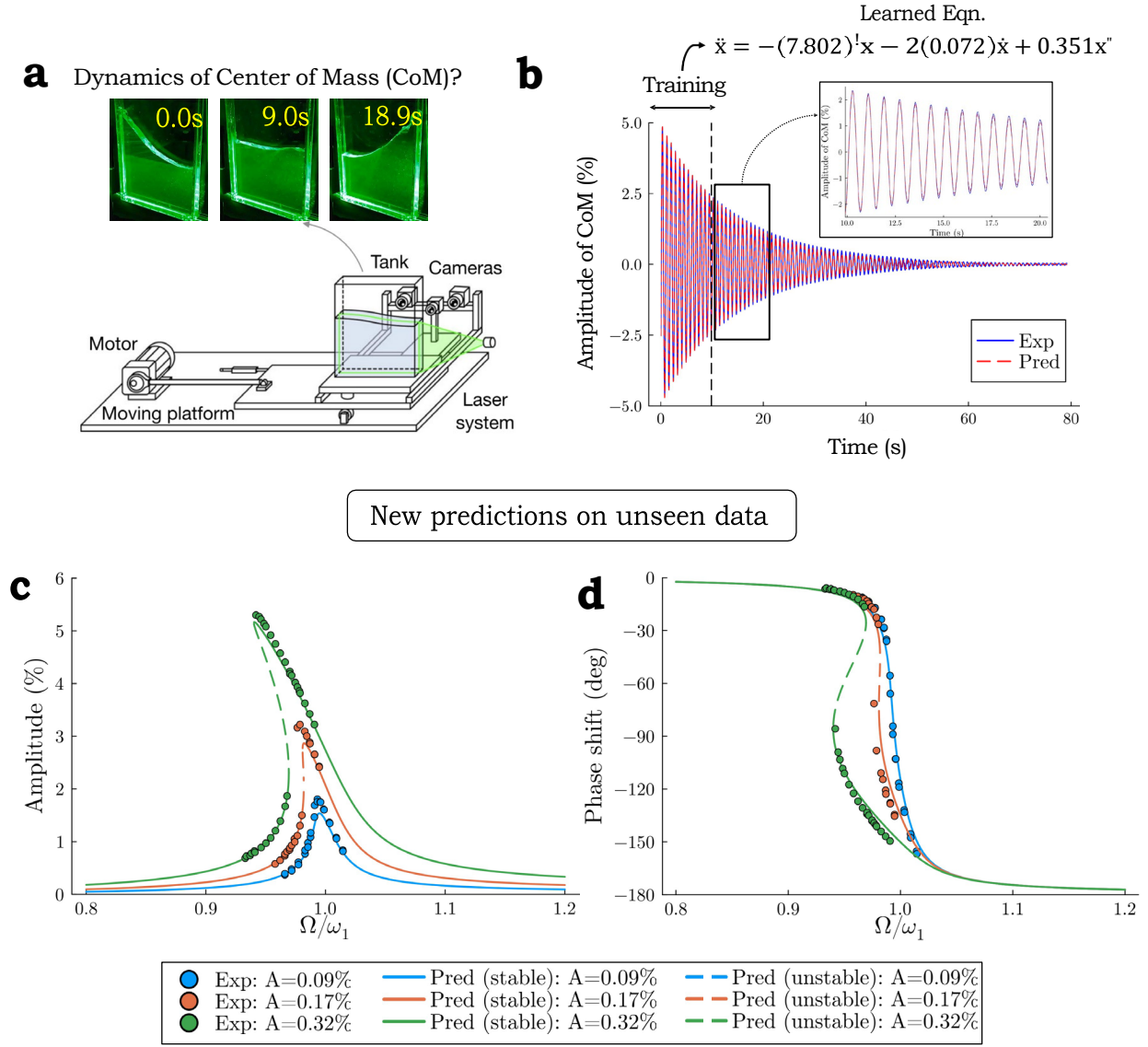

Figure 3: SLIC discovers center of mass (CoM) dynamics of a fluid in a sloshing tank and accurately predicts unseen forcing behavior. (a) Experimental design of sloshing tank experiments. We wish to determine the CoM dynamics extracted from images of unforced sloshing tank experiments. Image adapted from Bäuerlein and Avila, Cenedese et al.. (b) The governing equations extracted by SLIC using minimal training data from a single sloshing experiment accurately forecasts beyond training data. (c) Governing equations extracted by SLIC accurately predict CoM amplitude from unseen forcing conditions. The x-axis is forcing frequency relative to fundamental frequency. The y-axis is the steady-state CoM amplitude expressed as a percent of the tank width. (d) Governing equations extracted by SLIC accurately predict phase shift resulting from unseen forcing conditions. The x-axis is forcing frequency relative to fundamental frequency.

To demonstrate the ability of the algorithm to discover the functional form of a kernel from data, we chose to study a novel RNA-liposome cancer immunotherapy currently in clinical trials (NCT04573140). Interestingly, over time, heterogeneous aggregates form in this system, which are critical for therapeutic efficacy Mendez-Gomez et al. [b], Chung et al.. However, the underlying mechanism for this aggregation phenomenon is poorly understood, making this system an ideal candidate for testing the ability of SLIC to recover the kernel's functional form. Using DDMD permits us to extract the relevant kernel in a completely data-driven manner without manually fitting kernels.

Experimentally, we conduct quantitative fluorescence imaging of these particles (Fig. 4c) with single-cluster resolution as described in Ref. Chung et al.. From these images, we acquire time series data of the clusters, from which we can numerically estimate the kernel (See Methods). With the numerically estimated kernel, we can then use SLIC to extract the kernel's functional form from a candidate array of relevant kernel components (Fig. 4d) (See Methods for details). Each of the kernels in Fig. 4b can be reconstructed using an appropriate linear combination of the kernel components we train SLIC with, while permitting potential discovery of novel kernels or kernel combinations. Importantly, all relevant kernel functions (Fig. 4b) contain an unknown parameter, the fractal dimension ($d_f$), which quantifies the compactness of the aggregated clusters and determines important physical properties like density Elimelech et al.. Thus, in determining the kernel for this system, one must also estimate this unknown parameter simultaneously. To do so, we perform a round of model discovery with a fixed estimate for the fractal dimension, followed by an optimization loop that updates the unknown parameter; these steps are then iterated until convergence (Fig. S7a). Strikingly, from the candidate list of kernel components, SLIC identifies only the constant kernel as dominating the cluster-cluster interactions, which fits the experimental data quite well (Fig. S7b). Interestingly, this model with a constant kernel can be taken as an analytically solvable approximation to the Brownian kernel Elimelech et al., indicating that for this system, the cluster growth timescale is dictated by the diffusive nature of clusters (See Supporting Information for details).

This diffusion-dominated growth suggests a new strategy for the control of nanoparticle assembly in this system. Specifically, an initially lower concentration of particles should lead to a slower cluster growth rate. This hypothesis, inspired directly by the functional form of the kernel returned by the model, is indeed observed when we decrease the initial particle concentration (Fig. 4e). The linear growth rate of the average cluster size is also in quantitative agreement with the predicted kernel (See Supporting Information). Notably, other information criteria returns a kernel with unnecessary terms (Fig. S6b), thus obfuscating the dominant cluster-cluster interaction physics.

## 4 Conclusion

In this work, we present an algorithmic module that: (i) automatically and adaptively generates sparsity-enforcing parameters from the data and (ii) utilizes a novel sample-length-scaling information criterion, SLIC, to select models that are both accurate and simple. These parts work in synchrony to automatically generate and score models, which minimizes user-intervention for automating model discovery.

Our scoring criterion, SLIC, is designed with several objectives in mind. First, to respect the fact that if one assumes that there is an underlying ground truth model, then the optimal model should be invariant to the amount of data seen by an algorithm. Unlike existing information criteria utilized in DDMD, SLIC ensures that models do not exhibit a higher tendency to overfit with increasing data. Second, the functional form of SLIC determines that it equitably weighs accuracy and model complexity. Interestingly, in the sparse regression framework, minimizing the SLIC score also minimizes the likelihood that models are overfit due to the intimate relationship with variance minimization (see Supporting Text). This likely explains SLIC's performance, though it may be desirable to more directly tie SLIC to KLD minimization or a Bayesian approach Stoica and Selen.

Our results illustrate that SLIC with sparse regression correctly extracts governing equations from data spanning a broad spectrum of simulated ordinary and partial differential equations, including high noise levels and across a wide range of simulation conditions. Further, we demonstrated that the method can be used fruitfully with real experimental data.

This work demonstrates that careful consideration of the design of model-selection criteria is critical in DDMD. Though the form of our SLIC score does not currently possess hyperparameters, one could introduce parameters to tune the weighting of sparsity and goodness-of-fit (see Supporting Text for a discussion), which is analogous to the AIC-GIC relation Stoica and Selen, and may be suitable in other domains such as autoregression-based time-series forecasting, where model selection is critical Box et al. [2015].

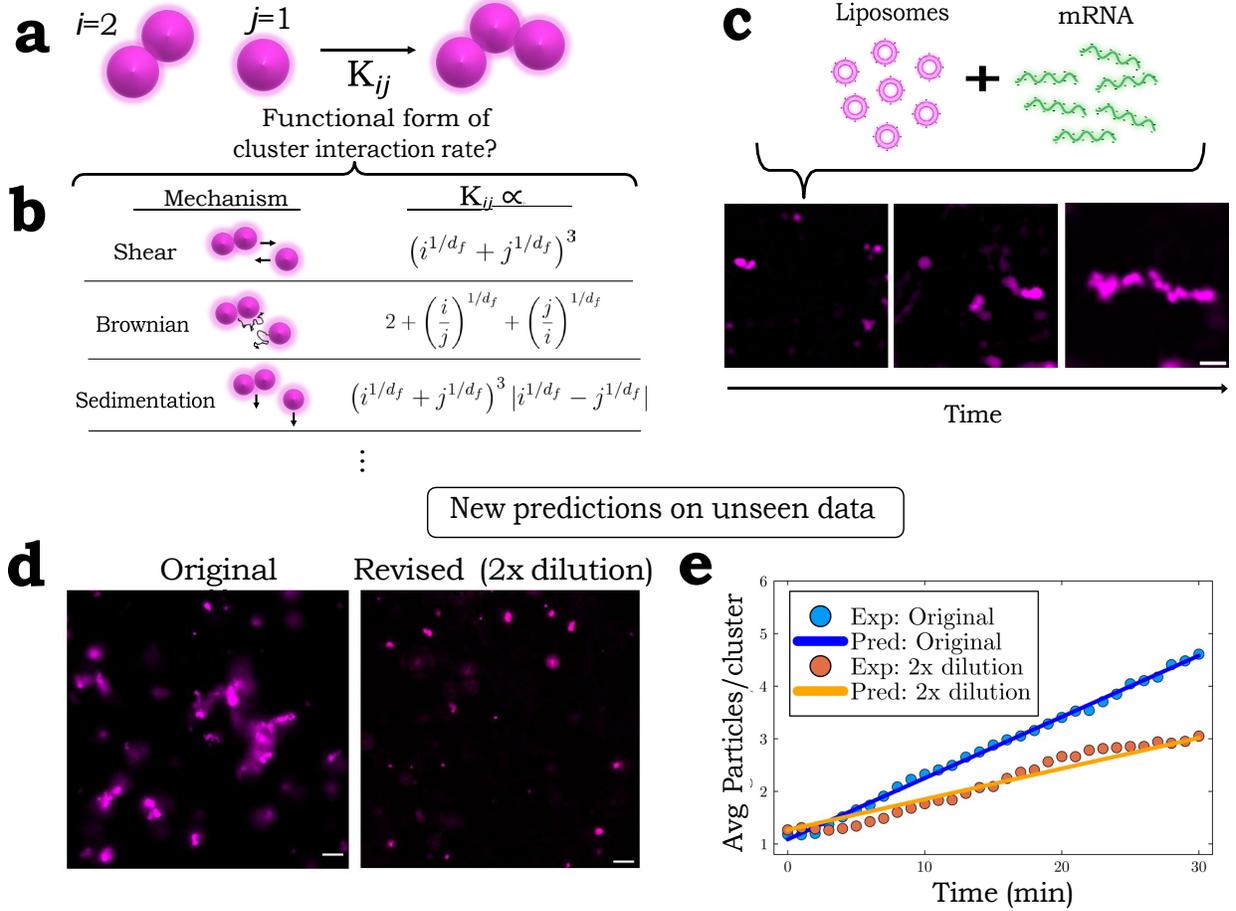

Figure 4: SLIC discovers the functional form of cluster-cluster interaction rate (the kernel) and predicts cluster growth rate of unseen initial conditions. (a) Illustration of the cluster-cluster interaction rate (i.e. the kernel, $K_{ij}$) for clusters of different particle number. The indices denote the number of monomers/particles per cluster. (b) Functional forms of kernels for different physical processes. Note that each depend on an unknown parameter, $d_f$. (c) Fluorescence timeseries images of the RNA-liposome system. Scale bar is 10m. (d) Images of the 'Original' and 'Revised' RNA-nanoparticle formulations. Original' denotes the undiluted solution. Scale bar is 5μm. (e) The constant kernel determined by SLIC accurately predicts the growth rate of 2x diluted solution of RNA-liposomes (See Methods).

## 5 Materials and Methods

### 5.1 Algorithm Performance Metrics

In this work, we use several metrics to quantify algorithm performance. Accuracy and false-positive rate (FPR) quantify the 'classification' performance of the algorithm (i.e., how often the algorithm correctly predicts the inclusion of a model term). With the shorthand TP = True Positive, FP = False Positive, TN = True Negative, FN = False Negative, these classification metrics are defined as follows:

$$\text{Accuracy} = \frac{\text{TP} + \text{TN}}{\text{TP} + \text{FP} + \text{TN} + \text{FN}} \tag{10}$$

$$\text{FPR} = \frac{\text{FP}}{\text{FP} + \text{TN}} \tag{11}$$

The accuracy reports how often the algorithm correctly predicts whether a model term should be included or not. The FPR reports how often a model term is incorrectly included in the returned model when it should not be. In the main text, the metrics are computed from averages of all the separate noise instantiations.

To quantify the algorithm's ability to recover the correct numerical values for each model, we utilized the scaled mean absolute error (SMAE), defined as:

$$\text{SMAE} = \frac{\|\Xi_{\text{true}} - \Xi_{\text{pred}}\|_1}{\|\Xi_{\text{true}}\|_1} \qquad (12)$$

Where $\|\cdot\|_1$ denotes the L1 norm, $\Xi_{\text{true}}$ is the true matrix of model coefficients, and $\Xi_{\text{pred}}$ is the predicted matrix of model coefficients returned by the algorithm.

To get statistics for each metric, at a fixed noise level, all systems studied were simulated 25 times with random instantiations of the noise, which is zero-mean Gaussian noise with unit magnitudes determined by

$$\% \text{ Noise} = \frac{100\sigma}{\text{std}(X)}$$

where $\sigma$ is the standard deviation of the Gaussian noise and std(X) is the state variable input data. The reported metrics at each noise percentage are the average of the respective metric across the different noise instantiations.

## 5.2 Information Criteria Used In Comparisons

In this section, we provide a list of the information criteria tested in this work. For each, $\hat{\epsilon}$ denotes the mean-square error of the model, n denotes the number of datapoints, and k denotes the number of free parameters (model complexity).

$$\text{AIC}(\hat{\epsilon}, k) = n\log(\hat{\epsilon}) + 2k$$
$$\text{AICc}(\hat{\epsilon}, k) = n\log(\hat{\epsilon}) + 2nk/(n - k - 1)$$
$$\text{HQIC}(\hat{\epsilon}, k) = n\log(\hat{\epsilon}) + k\log(\log(n))$$
$$\text{BIC}(\hat{\epsilon}, k) = n\log(\hat{\epsilon}) + k\log(n)$$
$$\text{KIC}(\hat{\epsilon}, k) = n\log(\hat{\epsilon}) - k\log(2\pi) + \log(|\Theta^T\Theta/\hat{\epsilon}|)$$
$$\text{BC}(\hat{\epsilon}, k) = n\log(\hat{\epsilon}) + n^{1/3}\left(1 + \frac{1}{2} + \ldots + \frac{1}{k}\right)$$
$$\text{SLIC}(\hat{\epsilon}, k) = n\log(\hat{\epsilon}k)$$

For the final term on the RHS of KIC, $|\cdot|$ denotes the determinant and $\Theta$ is the library matrix of nonzero terms; thus, it depends on both n and k.

## 5.3 The choice of candidate model terms

In this section, we provide a brief description of the libraries used in the main text. Specifics concerning the implementation can be found at https://github.com/MCHChung/SLIC.

### 5.3.1 Simulated systems

For Rossler, Brusselator, and Van der Pol we used a 3$^{\text{rd}}$ order polynomial basis in the state variables; this means that if (x, v) are the state variables, we consider $\{1, x, v, ..., x^3, x^2v, xv^2, v^3\}$. For Lorenz and Lotka-Volterra, we considered a 3$^{\text{rd}}$ order polynomial basis without constant term. For the nonlinear pendulum, we use a Fourier basis up to 5$^{\text{th}}$ order (i.e. $\{1, \sin(x), \cos(x), ..., \sin(5x), \cos(5x)\}$). For VdP, in which the library contains nonlinear functions of the velocity, we first estimate the velocity via a 4$^{\text{th}}$ order central difference method; this estimate and the position form the state variable inputs to the algorithm.

For Burger's equation, KdV, and KS the basis consisted of terms up to 4$^{\text{th}}$ order in spatial derivatives of the state variables, up to cubic terms in the state variables, and a nonlinear convective term (i.e. $u\partial_x u = 0.5\partial_x u^2$) in each state variable. For NLS, the basis consisted of terms up to 2$^{\text{nd}}$ order in spatial derivatives of the state variables, up to 3$^{\text{rd}}$ order polynomial basis in the state variables, and a nonlinear convective term in each state variable. For SG, the basis consisted of terms up to 2$^{\text{nd}}$ order in spatial derivatives of the state variables (in x and y), 1$^{\text{st}}$ order in time derivative, up to 2$^{\text{nd}}$ order polynomial basis in the state variables, and sin and cos trigonometric functions of the state variables.

### 5.3.2 Experimental systems

For the sloshing tank system, the derivative of the center of mass is estimated via a $4^{th}$ order central difference method. The center of mass data and its derivative estimate form the state variable inputs. For this system, we consider a $3^{rd}$ order polynomial basis (with no constant term, as that would imply a constant applied force) in these inputs. For the RNA-nanoparticle system, the basis consists of the kernel elements described in the text: 1, $i^{1/d_f}$, $j^{1/d_f}$, $(i/j)^{1/d_f}$, $(j/i)^{1/d_f}$, $(i^2)^{1/d_f}$, $(ij^2)^{1/d_f}$, $j^{3/d_f}$ and also $i^{1/d_f} + j^{1/d_f}$ multiplied by all $3^{rd}$ order terms, $i^{3/d_f}$, $(i^2j)^{1/d_f}$, $(ij^2)^{1/d_f}$, $j^{3/d_f}$. As described in the text, the indices i, j denote the number of monomers per cluster for the clusters interacting and $d_f$ is the unknown fractal dimension.

## 5.4 Data pre-processing

For all simulated systems, the workflow proceeds in several steps: (1) filter data to reduce noise (2) employ Galerkin projection to construct the library and targets and (3) carry out the hard-thresholding sparse regression algorithm with model selection using Algorithm 1 (or other information criteria). Below, we describe the first two steps.

### 5.4.1 Brief description of the fast Whittaker-Henderson filter

Often, the state variable data will be noisy. The Whittaker-Henderson filter Whittaker is one such approach for denoising that attempts to solve the following:

$$\arg\min_{\hat{y}} ||y - \hat{y}||_2^2 + s||D\hat{y}||_2^2 \quad (13)$$

Where y is our input data, $\hat{y}$ is the smoothed data to be determined, and $s||D\hat{y}||_2^2$ is a regularization penalty that penalizes non-smooth outputs, with s being the regularization strength and D a differential operator. However, the solution depends strongly on the choice of s. In this work, we implement a version of the filter that leverages the fast Fourier transform for rapid optimal determination of the smoothing parameter, s Garcia. This approach has the benefit of not only being fast, but easily generalizes to high-dimensional data, such as one might encounter with PDEs. The filtered input data, X, is then used to generate the library matrix $\Theta(X)$ and the targets, Y, if necessary. This is done for each simulated system.

### 5.4.2 Galerkin method for constructing reframed regression problem

To combat the challenges inherent in numerically estimating derivatives of state variables, which is a necessary component of performing DDMD with ODEs and PDEs, we use a modified version of a Galerkin method Messenger and Bortz [a,b]. This approach circumvents the need to estimate numerical derivatives of noisy data by modifying the underlying regression problem. Specifically, for the case of ODEs, if we assume that our targets are $Y = [y_1 \ldots y_m] = [\dot{x}_1 \ldots \dot{x}_m] = \dot{X}$, then, for each state variable, our regression problem takes the form:

$$\dot{x}_i = \sum_{j=1}^{l} \Xi_{ij}\, \theta_j(X) + \eta_i, \quad 1 \leq i \leq m \quad (14)$$

Where $\Xi_{ij}$ are the model coefficients to be determined. Now, the Galerkin method seeks to remove the necessity of computing numerical derivatives found on the left and right sides of this equality. Defining $\langle f \rangle_w = \int_{t_1}^{t_2} w(t) f(t)\, dt$ and multiplying by a known function w(t) on both sides of this equation, integrating by parts on the LHS, we obtain:

$$-\langle x_i \rangle_{\dot{w}} = \sum_{j=1}^{l} \Xi_{ij}\, \langle \theta_j(X) \rangle_w + \langle \eta_i \rangle_w, \quad 1 \leq i \leq m \quad (15)$$

Importantly, we have assumed that w(t) vanishes on the boundaries of the integration window. An infinite family of functions satisfies this property, and we utilize a symmetric, positive-valued polynomial function of the form

$$w(t) = w(t; c_1, t_1, t_2) = C_q\, [(t_2 - t)(t - t_1)]^q,$$

where q is a positive real number and $C_q$ is a normalizing coefficient chosen so that $\langle 1 \rangle_w = 1$. Thus, in short, we have transferred a numerical derivative of noisy data to the exact derivative of the known function w(t). In practice, the new, full-fledged Galerkin-projected vectors that serve as inputs to the sparse discovery algorithm are created by sliding our integration window along the input vector and concatenating the results of this integration.

For PDEs, the procedure is completely analogous, with the exception that the function we are multiplying by has multiple dimensions (i.e., $w = w(t, x_1, x_2, \ldots)$). For this we also use a symmetric polynomial kernel as described previously, with a single parameter. In one space and time dimension, this would be:

$$w(t, x; q, t_1, t_2, x_1, x_2) = C_q \left[(t_2 - t)(t - t_1)\right]^q \times C_q \left[(x_2 - x)(x - x_1)\right]^q,$$

Where $(x_1, x_2)$ is the boundary of integration of the x-dimension. We must also perform multidimensional integration. For all integration, we use trapezoidal integration provided by the Trapz.jl package in Julia.

To automatically determine the window size of integration, we repurpose the structural similarity index measure (SSIM) used in the image analysis field that was originally designed to determine the similarity between an image and a distortion of it Wang et al.. Such distortions may include convolutions or other filters. Given that the integration performed via the Galerkin method is a convolution, we reasoned that the SSIM can be used for automatically determining a window size of integration from an array of possible window sizes that minimally distorts the underlying data.

In summary, with this Galerkin method, our regression problem now takes the modified form:

$$\hat{\Xi} = \arg\min_{\Xi} \|\tilde{Y} - \tilde{\Theta}(X)\Xi\|_2^2 + \lambda^2 R(\Xi) \tag{16}$$

Where $\tilde{Y} = [-\langle \dot{x}_1 \rangle_w \ldots -\langle \dot{x}_m \rangle_w]$, $\tilde{\Theta}(X) = [\langle \theta_1(X) \rangle_w \ldots \langle \theta_l(X) \rangle_w]$, and $R(\Xi)$ is some regularization term to promote sparsity, with $\lambda$ the corresponding regularization weight parameter. It should be noted that it may be occasionally necessary to estimate numerical derivatives for libraries that contain nonlinear functions of the derivatives of state variables. In such cases, the Galerkin method is still of great use, as it still prevents successive, higher order estimates of derivatives for the target input.

### 5.5 Information regarding the simulated ODEs and PDEs

All data for ODEs featured in the main text were simulated using the OrdinaryDiffEq.jl package in Julia using the Tsit5 solver with a fixed time step of 0.01, except for Lorenz, Rössler, and Brusselator, which had a timestep of 0.001. For each ODE, three initial conditions were randomly chosen and simulated for the timespan shown for each system below at the aforementioned sampling frequency; this data was then concatenated and formed the input data. The data for all PDEs was acquired from references Messenger and Bortz [b], Rudy et al. and is provided in our GitHub for convenience. Further discussion of time stepping schemes may be found in those resources. Information pertinent to modifications of these systems is contained in the Supporting Information.

Below we give the equations studied in this work.

*Lorenz*

$$\frac{dx}{dt} = 10(y - x)$$
$$\frac{dy}{dt} = x(28 - z) - y$$
$$\frac{dz}{dt} = xy - \frac{8}{3}z$$

Timespan: $t \in [0, 10]$.

*Rossler*

$$\frac{dx}{dt} = -y - z$$
$$\frac{dy}{dt} = x + 0.2y$$
$$\frac{dz}{dt} = 0.5 + z(x - 5.7)$$

Timespan: $t \in [0, 10]$.

An information-based model selection criterion for data-driven model discovery    A PREPRINT*Lotka–Volterra*

$$\frac{dx}{dt} = x - 0.05xy$$
$$\frac{dy}{dt} = 0.05xy - y$$

Timespan: $t \in [0, 20]$.

*Nonlinear system*

$$\frac{dx}{dt} = 1 - 4x + x^2 y$$
$$\frac{dy}{dt} = 3x - x^2 y$$

Timespan: $t \in [0, 10]$.

*Van der Pol*

$$\frac{dx}{dt} = v$$
$$\frac{dv}{dt} = -x + 0.8(1 - x^2)v$$

Timespan: $t \in [0, 30]$.

*Nonlinear pendulum*

$$\frac{dx}{dt} = v$$
$$\frac{dv}{dt} = -4\sin(x)$$

Timespan: $t \in [0, 20]$.

*Burger's Equation*

$$\partial_t u(x, t) = -u(x, t)\partial_x u(x, t) - 0.1\partial_{xx} u(x, t)$$

Conditions: $u(x, 0) = e^{-(x+2)^2}$, $t \in [0, 10]$, $x \in [-8, 8]$. An evenly spaced grid of 101 timepoints and 256 spatial points was used.

*Kortweg-De Vries*

$$\partial_t u(x, t) = -6u(x, t)\partial_x u(x, t) - \partial_{xxx} u(x, t)$$

Conditions:

$$u(x, 0) = \frac{1^2}{2}\operatorname{sech}\left(\frac{\sqrt{1}}{2}(x + 20)\right) + \frac{0.5^2}{2}\operatorname{sech}\left(\frac{\sqrt{0.5}}{2}(x - L)\right), \quad t \in [0, 20], \ x \in [-30, 30].$$

An evenly spaced grid of 201 timepoints and 512 spatial points was used.

*Kuramoto–Sivashinsky*

$$\partial_t u(x, t) = -u(x, t)\partial_x u(x, t) - \partial_{xx} u(x, t) - \partial_{xxxx} u(x, t)$$

$$u(x, 0) = \cos\left(\frac{x}{1}\right)\left(1 + \sin\left(\frac{x}{1}\right)\right), \quad t \in [0, 150], \ x \in [0, 32\pi].$$

An evenly spaced grid of 1500 timepoints and 256 spatial points was used. Authors then subsampled 20% of the time points for a final resolution of $\Delta x = 0.393$ and $\Delta t = 0.5$.

*Nonlinear Schrödinger*

The full nonlinear Schrödinger equation, with $\psi$ a complex field, is:

$$i\partial_t \psi(x,t) = -\frac{1}{2}\partial_{xx}\psi(x,t) - |\psi|^2\psi(x,t)$$

Decomposed into real and imaginary components, $\psi = u + iv$:

$$\partial_t u(x,t) = -\frac{1}{2}\partial_{xx}v(x,t) - \left(u(x,t)^2 + v(x,t)^2\right)v(x,t)$$

$$\partial_t v(x,t) = \frac{1}{2}\partial_{xx}u(x,t) + \left(u(x,t)^2 + v(x,t)^2\right)u(x,t)$$

Conditions: Gaussian initial condition, $t \in [0, \pi]$, $x \in [-5, 5]$. An evenly spaced grid of 501 timepoints and 512 spatial points was used.

*Sine–Gordon*

$$\partial_{tt}u(x,y,t) = \partial_{xx}u(x,y,t) + \partial_{yy}u(x,y,t) - \sin(u(x,y,t))$$

Conditions:

$$u(x,y,0) = 2\pi e^{-8(x-0.5)^2 - 8y^2}, \quad t \in [0,5], \ x \in [-\pi, \pi], \ y \in [-1, 1].$$

### 5.6 RNA-Liposome kernel estimation and dual optimization problem for model discovery

We use the DiffEqFlux.jl package to estimate the kernel in a Smoluchowski model (Supporting Information) from our k-mer timeseries data that extends up to 50-mers. DiffEqFlux.jl determines the kernel that minimizes the $L_2$ error between the trajectories predicted by the Smoluchowski model and our data. Each element of this estimated kernel is then a function of a pair of k-mer indices, i, j. To encourage kernel symmetry (i.e., $K_{ij} = K_{ji}$), we concatenate this initial vector of estimated kernel values with its symmetric component; this forms the target matrix we are trying to find a model for. Once the kernel is estimated, for 1000 iterations we perform a sparse regression estimation of the model for each information criteria with a fixed $d_f$ followed by an optimization problem that updates $d_f$. The initial guess for $d_f$ is chosen randomly between 1.5 and 3, which are reasonable physical constraints bounding $d_f$. For the optimization loop that updates $d_f$, we minimize the loss function

$$\text{Loss} = \|\hat{K}_{\text{est}} - \Theta(i,j; d_f)\Xi\|_2^2$$

using the Optimization.jl package with the SAMIN() optimizer from the Optim.jl package that can perform bounded simulated annealing. This is repeated three times for each information criterion and the model with the best score for each criterion is chosen as the final model for evaluation.

## 6  Data availability, contributions, and funding acknowledgments

All data and code required to reproduce the results in this work can be found at https://github.com/MCHChung/SLIC

M.C.C. conceived the study. M.C.C. and A.Z. carried out the computations. All authors wrote the manuscript.

J.G. acknowledges National Institutes of Health grant R35GM146877 and Texas Global Faculty Research Seed Grant in support of this work

## References

Peter J. Schmid. Dynamic mode decomposition and its variants. 54:225–254. ISSN 0066-4189, 1545-4479. doi:10.1146/annurev-fluid-030121-015835. URL https://www.annualreviews.org/content/journals/10.1146/annurev-fluid-030121-015835. Publisher: Annual Reviews.

Marko Budišic´, Ryan Mohr, and Igor Mezic´. Applied koopmanisma). 22(4):047510. ISSN 1054-1500. doi:10.1063/1.4772195. URL https://doi.org/10.1063/1.4772195.

Dan Wilson. Data-driven identification of dynamical models using adaptive parameter sets. 32(2):023118. ISSN 1054-1500. doi:10.1063/5.0077447. URL https://doi.org/10.1063/5.0077447.

Matthew O. Williams, Ioannis G. Kevrekidis, and Clarence W. Rowley. A data–driven approximation of the koopman operator: Extending dynamic mode decomposition. 25(6):1307–1346. ISSN 1432-1467. doi:10.1007/s00332-015-9258-5. URL https://doi.org/10.1007/s00332-015-9258-5.

Miles Cranmer. Interpretable machine learning for science with PySR and SymbolicRegression.jl. URL http://arxiv.org/abs/2305.01582.

Patrick A. K. Reinbold, Logan M. Kageorge, Michael F. Schatz, and Roman O. Grigoriev. Robust learning from noisy, incomplete, high-dimensional experimental data via physically constrained symbolic regression. 12(1):3219. ISSN 2041-1723. doi:10.1038/s41467-021-23479-0. URL https://www.nature.com/articles/s41467-021-23479-0. Publisher: Nature Publishing Group.

Silviu-Marian Udrescu and Max Tegmark. AI feynman: A physics-inspired method for symbolic regression. 6(16):eaay2631. doi:10.1126/sciadv.aay2631. URL https://www.science.org/doi/10.1126/sciadv.aay2631. Publisher: American Association for the Advancement of Science.

Steven L. Brunton, Joshua L. Proctor, and J. Nathan Kutz. Discovering governing equations from data by sparse identification of nonlinear dynamical systems. 113(15):3932–3937. doi:10.1073/pnas.1517384113. URL https://www.pnas.org/doi/10.1073/pnas.1517384113. Publisher: Proceedings of the National Academy of Sciences.

Steven L. Brunton and J. Nathan Kutz. *Data-Driven Science and Engineering: Machine Learning, Dynamical Systems, and Control*. Cambridge University Press. doi:10.1017/9781108380690. URL https://www.cambridge.org/core/books/datadriven-science-and-engineering/77D52B171B60A496EAFE4DB662ADC36E.

Trevor Hastie, Robert Tibshirani, and Jerome Friedman. *The Elements of Statistical Learning*. Springer Series in Statistics. Springer. ISBN 978-0-387-84857-0 978-0-387-84858-7. doi:10.1007/978-0-387-84858-7. URL http://link.springer.com/10.1007/978-0-387-84858-7.

Kathleen Champion, Peng Zheng, Aleksandr Y. Aravkin, Steven L. Brunton, and J. Nathan Kutz. A unified sparse optimization framework to learn parsimonious physics-informed models from data. 8:169259–169271. ISSN 2169-3536. doi:10.1109/ACCESS.2020.3023625. URL https://ieeexplore.ieee.org/document/9194760/?arnumber=9194760. Conference Name: IEEE Access.

Daniel A. Messenger and David M. Bortz. Weak SINDy: Galerkin-based data-driven model selection. 19(3):1474–1497, a. ISSN 1540-3459. doi:10.1137/20M1343166. URL https://epubs.siam.org/doi/10.1137/20M1343166. Publisher: Society for Industrial and Applied Mathematics.

Daniel A. Messenger and David M. Bortz. Weak SINDy for partial differential equations. 443:110525, b. ISSN 0021-9991. doi:10.1016/j.jcp.2021.110525. URL https://www.sciencedirect.com/science/article/pii/S0021999121004204.

Kevin Egan, Weizhen Li, and Rui Carvalho. Automatically discovering ordinary differential equations from data with sparse regression. 7(1):1–10. ISSN 2399-3650. doi:10.1038/s42005-023-01516-2. URL https://www.nature.com/articles/s42005-023-01516-2. Publisher: Nature Publishing Group.

Pawan Goyal and Peter Benner. Discovery of nonlinear dynamical systems using a runge–kutta inspired dictionary-based sparse regression approach. 478(2262):20210883. doi:10.1098/rspa.2021.0883. URL https://royalsocietypublishing.org/doi/full/10.1098/rspa.2021.0883. Publisher: Royal Society.

Sheng Zhang and Guang Lin. Robust data-driven discovery of governing physical laws with error bars. 474(2217):20180305. doi:10.1098/rspa.2018.0305. URL https://royalsocietypublishing.org/doi/10.1098/rspa.2018.0305. Publisher: Royal Society.

Seth M. Hirsh, David A. Barajas-Solano, and J. Nathan Kutz. Sparsifying priors for bayesian uncertainty quantification in model discovery. 9(2):211823. doi:10.1098/rsos.211823. URL https://royalsocietypublishing.org/doi/10.1098/rsos.211823. Publisher: Royal Society.

Kevin Course and Prasanth B. Nair. State estimation of a physical system with unknown governing equations. 622(7982):261–267. ISSN 1476-4687. doi:10.1038/s41586-023-06574-8. URL https://www.nature.com/articles/s41586-023-06574-8. Publisher: Nature Publishing Group.

Jie Ding, Vahid Tarokh, and Yuhong Yang. Model selection techniques: An overview. 35(6):16–34, a. ISSN 1558-0792. doi:10.1109/MSP.2018.2867638. URL https://ieeexplore.ieee.org/document/8498082. Conference Name: IEEE Signal Processing Magazine.

P. Stoica and Y. Selen. Model-order selection: a review of information criterion rules. 21(4):36–47. ISSN 1558-0792. doi:10.1109/MSP.2004.1311138. URL https://ieeexplore.ieee.org/document/1311138. Conference Name: IEEE Signal Processing Magazine.

Hirotogu Akaike. Information theory and an extension of the maximum likelihood principle. In Emanuel Parzen, Kunio Tanabe, and Genshiro Kitagawa, editors, *Selected Papers of Hirotugu Akaike*, pages 199–213. Springer. ISBN 978-1-4612-1694-0. doi:10.1007/978-1-4612-1694-0_15. URL https://doi.org/10.1007/978-1-4612-1694-0_15.

Gideon Schwarz. Estimating the dimension of a model. 6(2):461–464. ISSN 0090-5364. URL https://www.jstor.org/stable/2958889. Publisher: Institute of Mathematical Statistics.

N. M. Mangan, J. N. Kutz, S. L. Brunton, and J. L. Proctor. Model selection for dynamical systems via sparse regression and information criteria. 473(2204):20170009. doi:10.1098/rspa.2017.0009. URL https://royalsocietypublishing.org/doi/full/10.1098/rspa.2017.0009. Publisher: Royal Society.

Alan A. Kaptanoglu, Lanyue Zhang, Zachary G. Nicolaou, Urban Fasel, and Steven L. Brunton. Benchmarking sparse system identification with low-dimensional chaos. 111(14):13143–13164. ISSN 1573-269X. doi:10.1007/s11071-023-08525-4. URL https://doi.org/10.1007/s11071-023-08525-4.

Nariaki Sugiura. Further analysis of the data by akaike's information criterion and the finite corrections. 7(1):13–26. ISSN 0361-0926. doi:10.1080/03610927808827599. URL https://doi.org/10.1080/03610927808827599. Publisher: Taylor & Francis.

E. J. Hannan and B. G. Quinn. The determination of the order of an autoregression. 41(2):190–195. ISSN 0035-9246. URL https://www.jstor.org/stable/2985032. Publisher: [Royal Statistical Society, Oxford University Press].

Rangasami L. Kashyap. Optimal choice of AR and MA parts in autoregressive moving average models. PAMI-4(2):99–104. ISSN 1939-3539. doi:10.1109/TPAMI.1982.4767213. URL https://ieeexplore.ieee.org/document/4767213. Conference Name: IEEE Transactions on Pattern Analysis and Machine Intelligence.

Jie Ding, Vahid Tarokh, and Yuhong Yang. Bridging AIC and BIC: A new criterion for autoregression. 64(6):4024–4043, b. ISSN 1557-9654. doi:10.1109/TIT.2017.2717599. URL https://ieeexplore.ieee.org/document/7953690. Conference Name: IEEE Transactions on Information Theory.

H. N. Abramson. The dynamic behavior of liquids in moving containers, with applications to space vehicle technology. URL https://ntrs.nasa.gov/citations/19670006555. NTRS Author Affiliations: ED. NTRS Report/Patent Number: NASA-SP-106 NTRS Document ID: 19670006555 NTRS Research Center: Legacy CDMS (CDMS).

A. E. P. Veldman, J. Gerrits, R. Luppes, J. A. Helder, and J. P. B. Vreeburg. The numerical simulation of liquid sloshing on board spacecraft. 224(1):82–99. ISSN 0021-9991. doi:10.1016/j.jcp.2006.12.020. URL https://www.sciencedirect.com/science/article/pii/S0021999106006139.

Carl Hubert. Behavior of spinning space vehicles with onboard liquids, 2nd edition, technical report b8030. URL https://ntrs.nasa.gov/citations/20160001550. NTRS Author Affiliations: Hubert Astronautics, Inc. NTRS Report/Patent Number: NASA/TP-2013-217917 NTRS Document ID: 20160001550 NTRS Research Center: Kennedy Space Center (KSC).

G Popov, S Sankar, T S Sankar, and G H Vatistas. Dynamics of liquid sloshing in horizontal cylindrical road containers. 207(6):399–406. ISSN 0954-4062. doi:10.1243/PIME_PROC_1993_207_147_02. URL https://doi.org/10.1243/PIME_PROC_1993_207_147_02. Publisher: IMECHE.

J. A. Romero, R. Hildebrand, M. Martinez, O. Ramirez, and J. A. Fortanell. Natural sloshing frequencies of liquid cargo in road tankers. 12(2):121–138. ISSN 1744-232X. doi:10.1504/IJHVS.2005.006379. URL https://www.inderscienceonline.com/doi/abs/10.1504/IJHVS.2005.006379. Publisher: Inderscience Publishers.

Bastian Bäuerlein and Kerstin Avila. Phase lag predicts nonlinear response maxima in liquid-sloshing experiments. 925:A22. ISSN 0022-1120, 1469-7645. doi:10.1017/jfm.2021.576. URL https://www.cambridge.org/core/journals/journal-of-fluid-mechanics/article/phase-lag-predicts-nonlinear-response-maxima-in-liquidsloshing-experiments/2CD68A3A9A3AFC648B311D6C01835988.

Mattia Cenedese, Joar Axås, Bastian Bäuerlein, Kerstin Avila, and George Haller. Data-driven modeling and prediction of non-linearizable dynamics via spectral submanifolds. 13(1):872. ISSN 2041-1723. doi:10.1038/s41467-022-28518-y. URL https://www.nature.com/articles/s41467-022-28518-y. Publisher: Nature Publishing Group.

Hector R. Mendez-Gomez, Anna DeVries, Paul Castillo, Christina von Roemeling, Sadeem Qdaisat, Brian D. Stover, Chao Xie, Frances Weidert, Chong Zhao, Rachel Moor, Ruixuan Liu, Dhruvkumar Soni, Elizabeth Ogando-Rivas, Jonathan Chardon-Robles, James McGuiness, Dingpeng Zhang, Michael C. Chung, Christiano Marconi, Stephen Michel, Arnav Barpujari, Gabriel W. Jobin, Nagheme Thomas, Xiaojie Ma, Yodarlynis Campaneria, Adam Grippin, Aida Karachi, Derek Li, Bikash Sahay, Leighton Elliott, Timothy P. Foster, Kirsten E. Coleman, Rowan J. Milner, W. Gregory Sawyer, John A. Ligon, Eugenio Simon, Brian Cleaver, Kristine Wynne, Marcia Hodik,

Annette M. Molinaro, Juan Guan, Patrick Kellish, Andria Doty, Ji-Hyun Lee, Tara Massini, Jesse L. Kresak, Jianping Huang, Eugene I. Hwang, Cassie Kline, Sheila Carrera-Justiz, Maryam Rahman, Sebastian Gatica, Sabine Mueller, Michael Prados, Ashley P. Ghiaseddin, Natalie L. Silver, Duane A. Mitchell, and Elias J. Sayour. RNA aggregates harness the danger response for potent cancer immunotherapy. 187(10):2521–2535.e21, a. ISSN 0092-8674, 1097-4172. doi:10.1016/j.cell.2024.04.003. URL https://www.cell.com/cell/abstract/S0092-8674(24) 00398-2. Publisher: Elsevier.

Shiza Malik, Khalid Muhammad, and Yasir Waheed. Nanotechnology: A revolution in modern industry. 28(2): 661. ISSN 1420-3049. doi:10.3390/molecules28020661. URL https://www.mdpi.com/1420-3049/28/2/661. Number: 2 Publisher: Multidisciplinary Digital Publishing Institute.

Michael J. Mitchell, Margaret M. Billingsley, Rebecca M. Haley, Marissa E. Wechsler, Nicholas A. Peppas, and Robert Langer. Engineering precision nanoparticles for drug delivery. 20(2):101–124. ISSN 1474-1784. doi:10.1038/s41573-020-0090-8. URL https://www.nature.com/articles/s41573-020-0090-8. Number: 2 Publisher: Nature Publishing Group.

Elvin Blanco, Haifa Shen, and Mauro Ferrari. Principles of nanoparticle design for overcoming biological barriers to drug delivery. 33(9):941–951. ISSN 1546-1696. doi:10.1038/nbt.3330. URL https://www.nature.com/articles/nbt.3330. Number: 9 Publisher: Nature Publishing Group.

Alexandre Albanese, Peter S. Tang, and Warren C.W. Chan. The effect of nanoparticle size, shape, and surface chemistry on biological systems. 14:1–16. ISSN 1545-4274. doi:https://doi.org/10.1146/annurev-bioeng-071811-150124. URL https://www.annualreviews.org/content/journals/10.1146/annurev-bioeng-071811-150124. Publisher: Annual Reviews Type: Journal Article.

Nazanin Hoshyar, Samantha Gray, Hongbin Han, and Gang Bao. The effect of nanoparticle size on in vivo pharmacokinetics and cellular interaction. 11(6):673–692. ISSN 1743-5889. doi:10.2217/nnm.16.5. URL https://www.futuremedicine.com/doi/10.2217/nnm.16.5. Publisher: Future Medicine.

Carlos L. Bassani, Greg van Anders, Uri Banin, Dmitry Baranov, Qian Chen, Marjolein Dijkstra, Michael S. Dimitriyev, Efi Efrati, Jordi Faraudo, Oleg Gang, Nicola Gaston, Ramin Golestanian, G. Ivan Guerrero-Garcia, Michael Gruenwald, Amir Haji-Akbari, Maria Ibáñez, Matthias Karg, Tobias Kraus, Byeongdu Lee, Reid C. Van Lehn, Robert J. Macfarlane, Bortolo M. Mognetti, Arash Nikoubashman, Saeed Osat, Oleg V. Prezhdo, Grant M. Rotskoff, Leonor Saiz, An-Chang Shi, Sara Skrabalak, Ivan I. Smalyukh, Mario Tagliazucchi, Dmitri V. Talapin, Alexei V. Tkachenko, Sergei Tretiak, David Vaknin, Asaph Widmer-Cooper, Gerard C. L. Wong, Xingchen Ye, Shan Zhou, Eran Rabani, Michael Engel, and Alex Travesset. Nanocrystal assemblies: Current advances and open problems. 18 (23):14791–14840. ISSN 1936-0851. doi:10.1021/acsnano.3c10201. URL https://doi.org/10.1021/acsnano.3c10201. Publisher: American Chemical Society.

Anish Rao, Sumit Roy, Vanshika Jain, and Pramod P. Pillai. Nanoparticle self-assembly: From design principles to complex matter to functional materials. 15(21):25248–25274. ISSN 1944-8244. doi:10.1021/acsami.2c05378. URL https://doi.org/10.1021/acsami.2c05378. Publisher: American Chemical Society.

M. Elimelech, J. Gregory, X. Jia, and R. A. Williams. Chapter 6 - modelling of aggregation processes. In M. Elimelech, J. Gregory, X. Jia, and R. A. Williams, editors, *Particle Deposition & Aggregation*, pages 157–202. Butterworth-Heinemann. ISBN 978-0-7506-7024-1. doi:10.1016/B978-075067024-1/50006-6. URL https://www.sciencedirect.com/science/article/pii/B9780750670241500066.

Hector R. Mendez-Gomez, Anna DeVries, Paul Castillo, Christina von Roemeling, Sadeem Qdaisat, Brian D. Stover, Chao Xie, Frances Weidert, Chong Zhao, Rachel Moor, Ruixuan Liu, Dhruvkumar Soni, Elizabeth Ogando-Rivas, Jonathan Chardon-Robles, James McGuiness, Dingpeng Zhang, Michael C. Chung, Christiano Marconi, Stephen Michel, Arnav Barpujari, Gabriel W. Jobin, Nagheme Thomas, Xiaojie Ma, Yodarlynis Campaneria, Adam Grippin, Aida Karachi, Derek Li, Bikash Sahay, Leighton Elliott, Timothy P. Foster, Kirsten E. Coleman, Rowan J. Milner, W. Gregory Sawyer, John A. Ligon, Eugenio Simon, Brian Cleaver, Kristine Wynne, Marcia Hodik, Annette M. Molinaro, Juan Guan, Patrick Kellish, Andria Doty, Ji-Hyun Lee, Tara Massini, Jesse L. Kresak, Jianping Huang, Eugene I. Hwang, Cassie Kline, Sheila Carrera-Justiz, Maryam Rahman, Sebastian Gatica, Sabine Mueller, Michael Prados, Ashley P. Ghiaseddin, Natalie L. Silver, Duane A. Mitchell, and Elias J. Sayour. RNA aggregates harness the danger response for potent cancer immunotherapy. 187(10):2521–2535.e21, b. ISSN 0092-8674, 1097-4172. doi:10.1016/j.cell.2024.04.003. URL https://www.cell.com/cell/abstract/S0092-8674(24) 00398-2. Publisher: Elsevier.

Michael C. Chung, Hector R. Mendez-Gomez, Dhruvkumar Soni, Reagan McGinley, Alen Zacharia, Jewel Ashbrook, Brian Stover, Adam J. Grippin, Elias J. Sayour, and Juan Guan. Multi-step assembly of an RNA-liposome nanoparticle formulation revealed by real-time, single-particle quantitative imaging. page 2414305. ISSN 2198-3844. doi:10.1002/advs.202414305. URL https://onlinelibrary.wiley.com/doi/abs/10.1002/advs.202414305. _eprint: https://onlinelibrary.wiley.com/doi/pdf/10.1002/advs.202414305.

George E.P. Box, Gwilym M. Jenkins, Gregory C. Reinsel, and Greta M. Ljung. *Time Series Analysis: Forecasting and Control, 5th Edition | Wiley*. Probability and Statistics. Wiley, 5 edition, 2015. ISBN 978-1-118-67502-1. URL https://www.wiley.com/en-us/Time+Series+Analysis%3A+Forecasting+and+Control%2C+5th+Edition-p-9781118675021.

E. T. Whittaker. On a new method of graduation. 41:63–75. ISSN 0013-0915, 1464-3839. doi:10.1017/S0013091500077853. URL https://www.cambridge.org/core/journals/proceedings-of-the-edinburgh-mathematical-society/article/on-a-new-method-of-graduation/744E6CBD93804DA4DF7CAC50507FA7BB.

Damien Garcia. Robust smoothing of gridded data in one and higher dimensions with missing values. 54(4):1167–1178. ISSN 0167-9473. doi:10.1016/j.csda.2009.09.020. URL https://www.sciencedirect.com/science/article/pii/S0167947309003491.

Z. Wang, E.P. Simoncelli, and A.C. Bovik. Multiscale structural similarity for image quality assessment. In *The Thrity-Seventh Asilomar Conference on Signals, Systems & Computers, 2003*, volume 2, pages 1398–1402 Vol.2. doi:10.1109/ACSSC.2003.1292216. URL https://ieeexplore.ieee.org/document/1292216.

Samuel H. Rudy, Steven L. Brunton, Joshua L. Proctor, and J. Nathan Kutz. Data-driven discovery of partial differential equations. 3(4):e1602614. doi:10.1126/sciadv.1602614. URL https://www.science.org/doi/10.1126/sciadv.1602614. Publisher: American Association for the Advancement of Science.